\documentclass[seceq]{ptptex}

\usepackage{graphicx}
\usepackage{wrapft}
\usepackage{color}

\title{A microscopic cluster-model study of $^3$He+$p$ scatterings}

\author{K. \textsc{Arai}$^{1,}$\footnote{E-mail: arai@nagaoka-ct.ac.jp}
S. \textsc{Aoyama}$^2$\footnote{E-mail: aoyama@cc.niigata-ct.ac.jp}, 
and Y. \textsc{Suzuki}$^3$\footnote{E-mail: suzuki@nt.niigata-ct.ac.jp}}

\inst{
$^1$Division of General Education, Nagaoka National College of
Technology, 888 Nishikatakai, Nagaoka,Niigata, 940-8532,Japan, \\
$^2$Integrated Information Processing Center,
Niigata University, Niigata 950-2181, Japan \\
$^3$Department of Physics, and Graduate School of Science and Technology,
Niigata University, Niigata 950-2181, Japan }

\abst{We calculate $^3$He+$p$ scattering phase shifts  
in two different microscopic cluster models,  
Model T and Model C,
in order to show the effects of tensor force 
as well as $D$-wave 
components in the cluster wave function.
Model T employs a realistic 
nucleon-nucleon potential and includes
the $D$-wave, whereas Model 
C employs an effective potential 
in which the tensor-force effect is 
considered to be renormalized into the central 
force and includes only the $S$-wave for the cluster intrinsic motion.
The $S$- and $P$-wave elastic scattering phase
shifts are obtained in the
\{$^3$He+$p$\}+\{$d$+2$p$\} coupled-channels calculation.
In  Model T, 
the $d$+2$p$ channel plays a significant role
in producing the $P$-wave resonant phase shifts but 
hardly affects the $S$-wave non-resonant phase 
shifts. In Model C, however, 
the effect of the $d$+2$p$
channel is suppressed in both of the $S$- and $P$-wave phase 
shifts, suggesting that it is  
renormalized mostly as the $^3$He(1/2$^+$)+$p$ channel in the 
resonance region.}

\begin{document}

\maketitle

\section{Introduction}

\par\noindent

For studies of structure and reactions of light nuclei, 
a cluster model is known to be one of 
successful models\cite{horiuchi86}.
A microscopic cluster model like  
the resonating group method (RGM)\cite{tang77,langanke94}
employs two ingredients.
The first is to assume, mainly 
because of its simplicity, that the nucleus is composed of 
a few $s$-shell clusters such as $\alpha$, $^3$H, $^3$He, and 
$d$ and that 
the antisymmetry requirement on the total wave 
function is properly taken into account.
The second is to employ an effective nucleon-nucleon ($N$-$N$)
interaction, e.g., 
the Minnesota potential (MN)\cite{thompson77}. 
The intrinsic wave function of the $s$-shell cluster 
is usually approximated with the $0s$ 
harmonic-oscillator (h.o.) function
whereas the cluster relative motion is solved accurately.
Corresponding to the simplified cluster wave functions, 
only the central, $LS$, and Coulomb terms 
of the $N$-$N$ interaction is usually employed, and the  
effects of the tensor force and the short-range 
repulsion which are present in a realistic interaction 
are assumed to be renormalized in 
the central force of the effective interaction. 
Though there are some calculations available which 
employ another type of effective $N$-$N$ interactions 
including the tensor force, 
its contribution was considered only 
for the cluster relative motion but not for 
the cluster intrinsic motion\cite{furutani79,csoto92,csoto93}.

In reality,
it is well known that the ground state of 
$^4$He, for example, has a large admixture 
of the $D$-wave component due to the tensor 
force, amounting to 
$P_d\approx$14\% for the AV8$^{\prime}$ 
potential\cite{pudliner07,kamada01,suzuki08}.
It is discussed that this $D$-wave component plays an important 
role for the $Q$-moment of the ground state of 
$^6$Li\cite{ryzhikh93,csoto92} as well as 
for the $\alpha$+$n$ $P$-wave phase shifts\cite{myo05}.
The $D$-wave component with  
($L$, $S$) = (2, 3/2) in the ground state of $^3$He is  
8.5\% for the 
AV8$^{\prime}$ potential and 7.0\% for the G3RS potential\cite{suzuki08}.
In order to understand more deeply 
the structure and reactions of light nuclei, 
it is important to test the microscopic cluster
model by taking into account both the 
tensor force of the $N$-$N$ 
interaction and the $D$-wave components in  
the $s$-shell clusters.

The purpose of the present article is to 
focus on the effect of the tensor force in 
$^3$He+$p$ $S$- and $P$-wave elastic scatterings by making   
comparative calculations in two different microscopic models, 
Model T and Model C.
In Model T, a realistic force 
including the tensor force 
is employed and the ($L$, $S$)=(2, 3/2) component  
of the $^3$He cluster as well as 
the $D$-wave component in the deuteron are taken 
into account. In
Model C, however, an effective potential 
without the tensor term
is employed and both the $^3$He and deuteron wave 
functions include only the $S$-wave components. 
Four $P$-wave broad resonances with spin and parity 2$^-$, 
1$^-$, 0$^-$, and 1$^-$ are observed in the low incident 
energy region of 4$-$7 MeV in the $^3$He+$p$ 
scattering\cite{tilley92}, but no resonant behavior 
is observed in the two $S$-wave phase shifts with 0$^+$ and 1$^+$. 
The $^3$He+$p$ scattering was previously investigated 
by various approaches with both realistic
\cite{deltuva07,pfitzinger01} and effective
interactions\cite{furutani79}.
Pfitzinger {\it et al.} calculated 
the elastic $^3$He+$p$ and $^3$H+$n$ scatterings
using the RGM with a realistic potential 
and discussed the phase shifts, analyzing powers and 
cross sections~\cite{pfitzinger01}. 
In the present article, 
we will clarify both the effects of the $D$-wave
components in the cluster wave functions 
and the mixing of the $d$+2$p$ channel by comparing 
the results of Models T and C.

Firstly we calculate the phase shifts in Model T. 
In this calculation, the $d$+2$p$ channel 
as well as other spin-parity states of $^3$He 
up to 5/2$^{\pm}$ 
are included in order to take into account the breakup or distortion
effect of $^3$He. 
These states of $^3$He other than 
the 1/2$^+$ state are actually continuum states but they 
are approximated with discretized states in the present calculation.
Secondly we repeat the phase shift calculation 
in Model C using the interaction which contains no 
tensor force.

The organization of this article is as follows. 
In the next section, Model T 
as well as Model C are briefly explained.
In Sect.3, the phase shifts obtained by both models 
are presented. Calculations in the similar 
models are also performed for the ground state of $^4$He and 
$^3$H+$p$ $S$-wave scattering phase shift. 
Summary is given in Sect. 4. 

\section{Model}
\label{model}

In the present study, we have employed 
the microscopic cluster model as formulated by 
the RGM\cite{tang77,langanke94}.
In this method, all the nucleons are treated explicitly
and they are assumed to be arranged in several clusters. 
The wave function consisting of two clusters ($A$+$B$) is given as
\begin{equation}
\Psi^{JM\pi}_{AB} = \sum_{i=1}^{N_A}  \sum_{j=1}^{N_B}
{\cal A}\Big\{
[[\Phi^{A, i}_{I_A, \pi_A}\Phi^{B, j}_{I_B, \pi_B}]_I \; 
\chi^{A,i,B,j}_{\ell}
(\mbox{\boldmath$\rho$})]_{JM} \Big\},
\label{eqn:a1}
\end{equation}
where $\Phi^{A, i}_{I_A, \pi_A}$ and $\Phi^{B, j}_{I_B, \pi_B}$ are the
intrinsic wave functions of the clusters $A$ and $B$, 
and their spins, $I_A$ and $I_B$, are coupled to 
the channel spin $I$ as indicated by the 
square bracket [\ \ ]. 
The symbol $N_A$($N_B$) stands for the number
of the basis set for the cluster intrinsic wave function of
the cluster $A$($B$). 
The first state with $i$($j$)=1 is the ground state 
and the states with $i\ge 2$ denote pseudostates. 
The ground states of $^3$He and $d$ are bound
but those of the $2p$ and $d$(0$^+$) clusters are virtual states.
Here $2p$ stands for a di-proton cluster.   
The cluster relative motion function 
$\chi^{A,i,B,j}_{\ell}(\mbox{\boldmath$\rho$})$ with
the partial wave $\ell$ is specified by 
the cluster relative distance coordinate $\mbox{\boldmath$\rho$}$.
The total wave function (\ref{eqn:a1}) is properly antisymmetrized 
as indicated by 
the intercluster antisymmetrizer ${\cal A}$. It contains  
no center-of-mass wave function, and has 
good total angular momentum $JM$ and parity $\pi$.

We take into account not only the  $^3$He+$p$ elastic 
channel but also the inelastic channels including 
the different spin-parity states of $^3$He  
and the rearrangement channel of $d$+2$p$ as well. 
For all of the $^3$He, $d$, and $2p$
clusters in each spin parity states, the 
pseudostates are taken into account in the present calculation.
The pseudostates, when included in the phase-shift 
calculation, are expected to take account of the distortion of 
the clusters of the entrance channel.\cite{beck84,kanada85} 
In the case of the coupled-channels calculation of 
\{$A$+$B$\}+\{$A'$+$B'$\}+$\cdots$,
the total wave function reads 
\begin{equation}
\Psi^{JM\pi} = \Psi^{JM\pi}_{AB} + \Psi^{JM\pi}_{A'B'} + \cdots  \,\, .
\label{eqn:a2}
\end{equation}

The intrinsic wave functions of $^3$He 
used in Eq.~(\ref{eqn:a1})
are given by three-body calculations 
of $p$+$p$+$n$ as 
($\alpha$ stands for $^3$He) 
\begin{equation}
\Phi^{\alpha,i}_{I_\alpha, M_\alpha, \pi_\alpha} =
\sum^{N_{\alpha}}_{\lambda_{\alpha}=1} C^i_{\lambda_{\alpha}}
{\cal A}\Big\{\left[ \phi_{S_\alpha T_\alpha} \left[
\Gamma_{\ell_1}(\nu_1,\mbox{\boldmath$\rho$}_1)
\Gamma_{\ell_2}(\nu_2,\mbox{\boldmath$\rho$}_2)
\right]_{L_{\alpha}} \right]_{I_{\alpha}, M_{\alpha}} \Big\}.
\label{eqn:a3}
\end{equation}
The subscript $\lambda_{\alpha}$ stands for a set of the labels
$\{S_{\alpha},T_{\alpha},L_{\alpha},\ell_1,\ell_2,\nu_1,\nu_2\}$ and
$C^i_{\lambda_{\alpha}}$ is the coefficients of the $i$'th
eigenvalue obtained by diagonalizing the $^3$He cluster
intrinsic Hamiltonian.
The Gaussian basis function 
$\Gamma_{\ell_i}(\nu_i, \mbox{\boldmath$\rho$}_i)$
are given in Eqs. (4) and (5) of Ref.~\citen{arai01} and 
$\mbox{\boldmath$\rho$}_1$, $\mbox{\boldmath$\rho$}_2$
are the Jacobi coordinates in the $p$+$p$+$n$ system 
with $\ell_1$, $\ell_2$ 
denoting the corresponding orbital angular momenta.  
The function $\phi_{S_{\alpha} T_{\alpha}}$ is 
the spin and isospin part of $^3$He with  
$S_{\alpha}$ and $T_{\alpha}$ being the total spin and total isospin,
respectively.
The total angular momentum and 
parity ($I_{\alpha}^{\pi_{\alpha}}$) of $^3$He is 
taken into account up to 5/2$^{\pm}$  with the restriction 
of $\ell_1$, $\ell_2$ $\leq$ 2
and $S_{\alpha}$=1/2 or 3/2 and $T_{\alpha}$=1/2. 
The wave function for the deuteron cluster, 
denoted as $\beta$, has a form similar to Eq.~(\ref{eqn:a3}) as
\begin{equation}
\Phi^{\beta, j}_{I_{\beta}, M_{\beta}, \pi_{\beta}} =
\sum_{\lambda_{\beta=1}}^{N_{\beta}} C^j_{\lambda_{\beta}}
{\cal A}\Big\{\left[ \phi_{S_{\beta} T_{\beta}}
\Gamma_{L}(\nu_1,\mbox{\boldmath$\rho$}_1)
\right]_{I_{\beta}, M_{\beta}} \Big\}.
\label{eqn:a4}
\end{equation}
The subscript $\lambda_{\beta}$ stands for a set of the labels
$\{S_{\beta}, T_{\beta}, L, \nu_1\}$.
For the deuteron, $I_{\beta}^{\pi_{\beta}}$=1$^+$ with 
$S_{\beta}$=1, $T_{\beta}$=0, and $L$=0 or 2. 
We also consider the $pn$ cluster  
which have $I_{\beta}^{\pi}$=0$^+$ ( 
$S_{\beta}$=0, $T_{\beta}$=1, $L$=0).
The 2$p$ cluster are given similarly to  
Eq.~(\ref{eqn:a4}) with $I_{\beta}^{\pi_{\beta}}$=0$^+$ 
($S_{\beta}$=0, $T_{\beta}$=1, $L$=0).
The spatial parts of the cluster 
wave functions are given in terms of a combination of 
Gaussian basis functions with different values of 
$\nu_i$.

The wave functions given in 
Eq.~(\ref{eqn:a1})$\sim$(\ref{eqn:a4}) 
are obtained by solving the respective $A$-nucleon 
Schr$\ddot{\rm o}$dinger
equations with the Hamiltonian 
\begin{equation}
{H} = \sum_{i=1}^A {T}_i -{T}_{CM} + \sum_{i<j}^A {V}_{ij} ,
\label{eqn:a5}
\end{equation}
where ${T}_i$ is the kinetic energy of 
the $i$th nucleon,
${T}_{CM}$ is the kinetic energy of the center-of-mass 
motion,
and ${V}_{ij}$ is the nucleon-nucleon interaction.

The cluster relative motion $\chi^{A,i,B,j}_{\ell}(\mbox{\boldmath$\rho$})$
in Eq.(\ref{eqn:a1}) is solved with 
the microscopic $R$-matrix 
method(MRM),\cite{baye77} in which the configuration 
space for the relative motion between the clusters is divided 
into two regions, inner and outer, by a channel radius.  
The relative wave function in the inner region   
is approximated with a superposition
of Gaussian basis functions 
$\Gamma_{\ell}(\nu, \mbox{\boldmath$\rho$})$
with various range parameters $\nu$.\cite{arai01} 
The same set of Gaussian basis functions
is employed for all the channels.
The range parameters are taken in the range of  
0.1fm$<$ $b(=1/\sqrt{\nu})$ $<$15fm, and the 
number of $\nu$ is 13 for Model T
and 15 for Model C in the $^3$He+$p$
phase-shift calculation.
The channel radius is chosen as 13.5fm for Model T 
and 15fm for Model C.
In the case of $^3$H+$p$ scattering 
the number of basis set is 15 and the channel radius is
taken as 15fm for both the models of T and C. 
In order to avoid the numerical instability in the MRM calculation, 
the range parameters in Eqs.~(\ref{eqn:a3}) and (\ref{eqn:a4}) 
for the cluster intrinsic motion 
are taken in the range of $b_i(=1/\sqrt{\nu_i})$ $<$5fm.
The relative wave function in 
the inner region is connected, at the channel radius, 
smoothly to the asymptotic form of the relative 
wave function which is expressed in 
terms of the Coulomb functions and the scattering $S$-matrix to be 
determined.

In Model T, we employ the G3RS potential
which reproduces the $N$-$N$ scattering data 
reasonably well\cite{tamagaki68}.
We use this 3-range Gaussian potential mainly because 
it saves a computer time compared to  more recent potentials  
such as AV8$^{\prime}$ potential\cite{pudliner07}. 
The central, $LS$ and 
tensor terms of the G3RS potential are included together 
with the Coulomb potential, but the $L^2$ and $(LS)^2$ terms 
which give a negligible contribution
are omitted.  
The $p$+$p$+$n$ three-body wave function 
for $^3$He is obtained by taking into 
account the partial waves up to the $D$-wave for each 
Jacobi coordinate.

\begin{table}[t]
\caption{
Energies $E$ (MeV) and rms radii $r$ (fm) 
and probabilities of the $L$=2 component $P_D$(\%) 
of the $^3$He and the deuteron. 
`Limited' denotes the results obtained 
with a limited number of basis set as explained in text, while 
`Full' the converged results~\cite{suzuki08} 
obtained with large enough basis set. 
The experimental energy of $^3$He is $-$7.718MeV\cite{tilley92}.}
\label{table:1}
\begin{center}
\begin{tabular}{ccccccc} \hline \hline
   & Wave function   &       & \multicolumn{2}{c}{G3RS} &
\multicolumn{2}{c}{MN} \\ 
   &    &       &  Limited & Full & Limited & Full \\ \hline
$^3$He & 0s h.o. &  $E$  &   &   & $-$5.28  &  \\
       &         &  $r$  &   &   & 1.71     &     \\
$^3$He & $p$+$p$+$n$ &  $E$  & $-$6.55  & $-$7.08  & $-$7.70  & $-$7.71 \\
       &             &  $r$  & 1.76     & 1.82     & 1.73     & 1.74    \\
       &             &  $P_D$  & 6.2 & 7.0  & -- & -- \\
$d$    & $p$+$n$     &  $E$  &  $-$2.09  & $-$2.28  &  $-$2.10    & $-$2.20  \\
       &             &  $r$  &  1.71     & 1.98  &  1.63 &  1.95  \\
       &             &  $P_D$  & 5.0 & 4.8  &--  & -- 
 \\  \hline
\end{tabular}
\end{center}
\end{table}

Model T takes into account the configurations of 
\{$^3$He(1/2$^\pm$, 3/2$^{\pm}$, 5/2$^{\pm}$)+$p$\}
+ \{$d$(0$^+$,1$^+$)+2$p$(0$^+$)\}
and the partial wave $\ell$ in Eq.~(\ref{eqn:a1}) 
up to $\ell$=3.
The 1/2$^+$ ground state of $^3$He is approximated with 
fifteen Gaussian basis functions which are 
selected by the stochastic variational method (SVM)
\cite{kukulin77,varga94}.
Table ~\ref{table:1} lists 
the energy and root mean square (rms) radius 
with this limited number of basis set 
as well as the convergent result.\cite{suzuki08} 
This restriction for the basis dimension is necessary in 
order to make the $^3$He+$p$ calculation feasible.
All of the other spin-parity states of $^3$He are 
unbound and they are approximated with the 
wave functions of a bound-state type  using 
ten basis functions which are selected randomly.
For the other spin-parity states of $^3$He,
we test a different number of basis set 
in order to check the sensitivity of the
basis choice to the 0$^+$ and 0$^-$ phase shifts: 
The number of basis set is 22, 18, 14, 20, and 16 for 
the 3/2$^+$, 5/2$^+$, 1/2$^-$, 3/2$^-$, and 5/2$^-$ states,
respectively. 
Both phase shifts are hardly affected by 
these choices  
when the $d$+2$p$ channel is added in the calculation.
Without the $d$+2$p$ channel, the difference in the 0$^-$ phase shift 
is found to be less than 5 degree in 
the energy range $E_{\rm cm}\leq$10\,MeV, 
while the 0$^+$ phase shift remains unchanged.

The deuteron wave function 
in Model T is given by a superposition of three $S$-wave
plus three $D$-wave Gaussian basis functions with  
$\nu_1$ in Eq.~(\ref{eqn:a4}) being  
2.40, 0.266, 0.0400 fm$^{-2}$ for the $S$-wave and 
0.974, 0.328, 0.0930 fm$^{-2}$ for the $D$-wave.
Energy and rms radius of the deuteron 
with this limited basis set as well as 
those with a larger basis set are 
compared in Table~\ref{table:1}.
Since the rms radius of the deuteron with 
the limited basis 
is smaller than the one with the full set,
one might think that the contribution of the $d$+2$p$
channel is underestimated in the present $^3$He+$p$ calculation. 
As will be discussed in Sect.3.3,
these limited wave functions of $^3$He and the deuteron
are found to be useful in accounting for the binding 
energy of the 0$^+$ ground state of $^4$He within only 0.88MeV, 
compared with a more precise SVM calculation, which indicates 
that the pseudostates of $^3$He and $d$ taken into account 
in the present calculation make it possible to describe 
the distortion of the clusters in the $^3$He+$p$ channel.

The three $S$-wave bases used for the deuteron wave function 
are also used to describe 
the unbound 0$^+$ states in the $pn$ and $2p$ cluster  
systems by the bound state-type wave functions.
Some possible effects of the 
three- and four-body channels of $d$+$p$+$p$ and $p$+$p$+$p$+$n$ 
are expected to be included in the the present $R$-matrix method 
through the $d$(0$^+$, 1$^+$)+2$p$ two-cluster channels. 
This sort of approximations was employed to discuss 
the three-body resonances in $^9$Be and $^{12}$C
and gave the results consistent 
with those of the three-body complex scaling 
method\cite{arai03}.

\begin{figure}[t]
\centerline{\includegraphics[width=10 cm,height=8 cm]{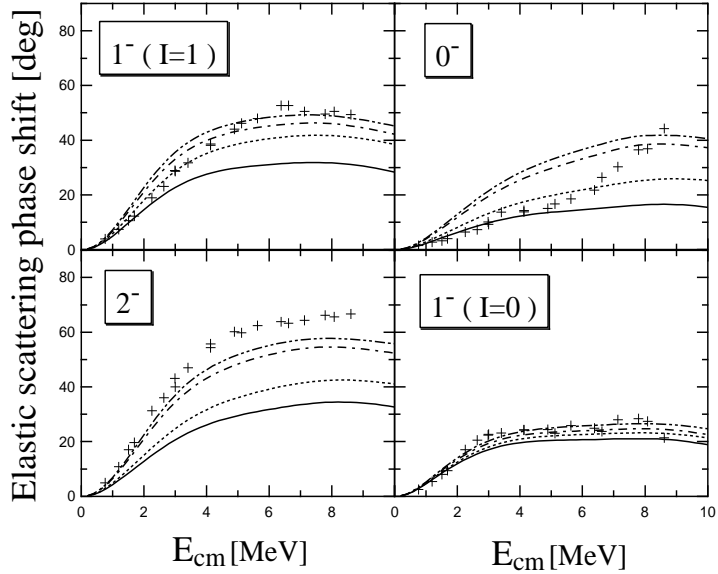}}
\caption{The $^3$He+$p$ $P$-wave elastic scattering phase shifts
calculated by Model T (G3RS potential). 
In Model T, we employ the G3RS potential
The channel spin $I$ is 0 or 1 for the $1^-$ state, 
and 1 for the $0^-$ and $2^-$ 
states. The lines denote the results obtained  
including the following configurations: 
Solid  $^3$He(1/2$^+$)+$p$; 
\, Dotted $^3$He(1/2$^{\pm}$, 3/2$^{\pm}$, 5/2$^{\pm}$)+$p$; 
\, Dash-dotted \{$^3$He(1/2$^+$)+$p$\} + \{$d$(0$^+$,1$^+$)+2$p$(0$^+$)\}; 
\, Dash-double-dotted 
\{$^3$He(1/2$^{\pm}$, 3/2$^{\pm}$, 5/2$^{\pm}$)+$p$\} + 
\{$d$(0$^+$,1$^+$)+2$p$(0$^+$)\}. 
The crosses denote the experimental data\cite{tombrello65} and
the error bars of the data are omitted.}
\label{fig:g3rsa}
\end{figure}

In Model C, the MN potential\cite{thompson77} 
with $u$=0.98  
is employed as the effective $N$-$N$ interaction.
This interaction 
can reproduce the $n$-$p$ triplet and $p$-$p$ singlet $S$-wave
scattering lengths and effective ranges.
With the spin-orbit term of Reichstein and Tang
(set IV)\cite{reichstein70},
this potential can reproduce low-energy 
$\alpha$+$n$ phase shifts for the $S$- and $P$-waves\cite{csoto93a}.
Since the MN potential can fairly well reproduce 
the binding energies of $d$, $^3$H, and $^4$He 
without a tensor term\cite{suzuki08}, 
any additional tensor term makes 
these $s$-shell clusters seriously overbound.  
Therefore, the MN potential without any tensor term 
is employed in Model C. Here 
the configurations of \{$^3$He(1/2$^+$)+$p$\}+
\{$d$(0$^+$, 1$^+$)+2$p$(0$^+$)\} are included 
and  the wave functions of $^3$He and $d$
contain only $S$-wave components. 

For the $^3$He cluster,
we employ two different wave functions.
One is a superposition of
the $(0s)^3$ h.o. functions with four different 
oscillator parameters,
$\nu$=1.234, 0.548, 0.208, 0.0696 fm$^{-2}$.
Second is the wave function obtained in the 
$p$+$p$+$n$ three-body calculation 
in which the partial wave for each Jacobi coordinate
is restricted to the $S$-wave only. The wave function is 
a combination of fifteen Gaussians selected by the SVM.
The deuteron wave function is given by four Gaussian basis set 
where the Gaussian parameters 
$\nu_1$ in Eq.~(\ref{eqn:a4}) are 
1.297, 0.552, 0.198, 0.040fm$^{-2}$.
The 0$^+$ states of 
the $pn$ and $pp$ clusters 
are approximated with the bound-state 
type wave functions 
using the same basis set as used in the deuteron wave function.
Energies and rms radii of $^3$He and $d$ are listed in Table~\ref{table:1}. 
The partial waves for the cluster relative motion are taken up to 
$\ell$=3.

\section{Results}

\subsection{$^3$He+$p$ elastic scattering in Model T}

\begin{figure}[t]
\centerline{\includegraphics[width=12 cm,height=5 cm]{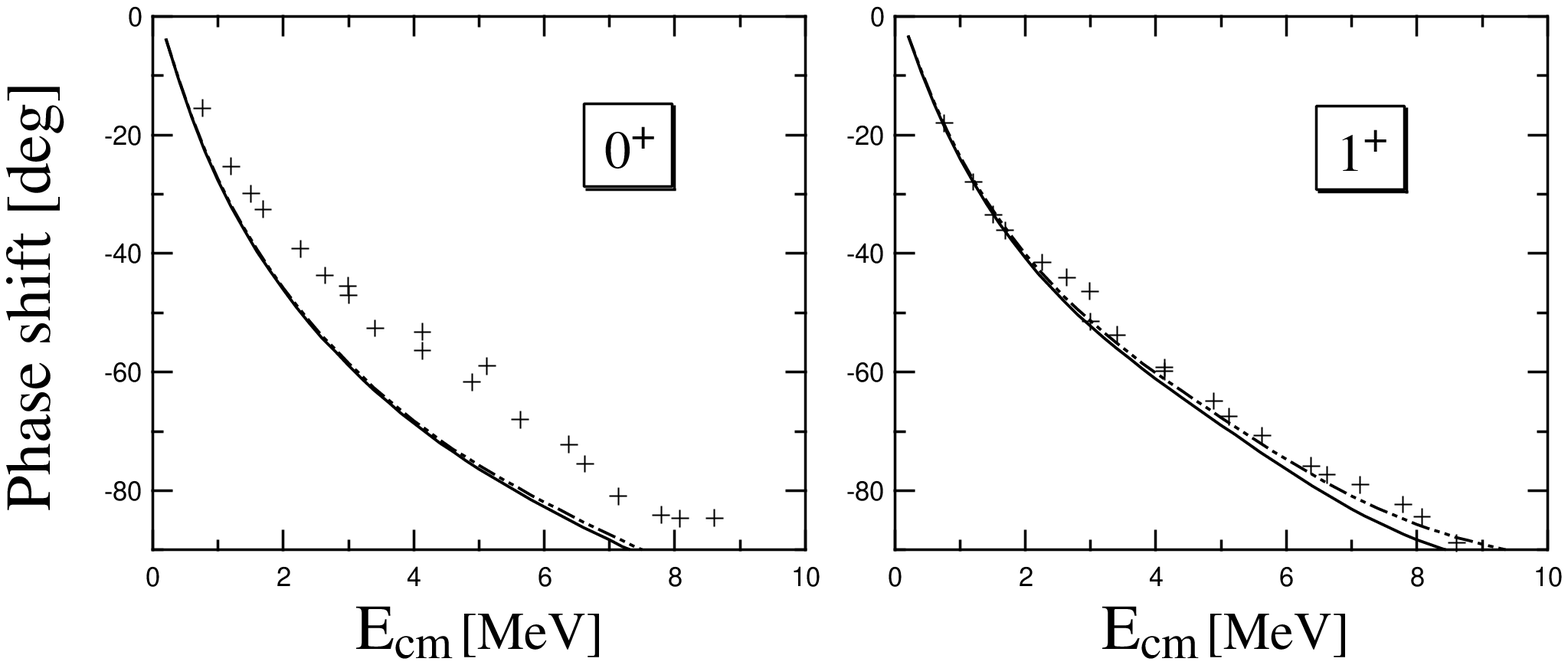}}
\caption{The $^3$He+$p$ $S$-wave elastic scattering phase shifts
calculated by Model T (G3RS potential).
The solid and dash-double-dotted lines denote the results with 
the $^3$He(1/2$^+$)+$p$ and
\{$^3$He(1/2$^{\pm}$, 3/2$^{\pm}$, 5/2$^{\pm}$)+$p$\} 
+ \{$d$(0$^+$,1$^+$)+2$p$(0$^+$)\} configurations, respectively.
The crosses denote the experimental data\cite{tombrello65}.}
\label{fig:g3rsb}
\end{figure}

Figure~\ref{fig:g3rsa} displays the $^3$He+$p$ $P$-wave 
elastic scattering phase shifts  in comparison 
with experiment.\cite{tombrello65} 
These are obtained  in Model T with the 
inclusion of the $D$-wave for the $^3$He cluster.
As mentioned in the Introduction, the $P$-wave scattering 
produces four negative-parity states with 
2$^-$, 1$^-$($I$=0, 1), and 0$^-$, all of which correspond to  
the low-lying broad resonances of $^4$Li. 
The solid, dotted, dash-dotted, and dash-double-dotted lines 
of the figure denote the results obtained by including the 
configurations of 
$^3$He(1/2$^+$)+$p$, $^3$He(1/2$^{\pm}$, 3/2$^{\pm}$, 5/2$^{\pm}$)+$p$, 
\{$^3$He(1/2$^+$)+$p$\}+\{$d$+2$p$\}, and 
\{$^3$He(1/2$^{\pm}$, 3/2$^{\pm}$,
5/2$^{\pm}$)+$p$\}+\{$d$+2$p$\}, respectively, 
where the spin-parity of the $d$ cluster 
includes both 0$^+$ and 1$^+$, and that  
of the $2p$ cluster is 0$^+$. 
The calculation with all the configurations 
(dash-double-dotted lines) give results similar to those of the RGM calculation\cite{pfitzinger01}. 
Our result for the 0$^-$ state is in 
disagreement with the experimental data in 
the whole energy range. A further 
consideration for the model space or the nucleon-nucleon interaction 
such as three-body forces 
may be necessary 
in order to reproduce the experimental phase shifts.

Apparently, the single-channel calculation of $^3$He(1/2$^+$)+$p$ is 
quite insufficient for reproducing 
the $P$-wave phase shifts
except for the 1$^-$($I$=0) case. 
The $d$(1$^+$)+2$p$ channel($I$=1) hardly changes 
the 1$^-$($I$=0) phase shift whereas it gives a significant 
contribution to the other $P$-wave phase shifts with $I$=1. 
The $d$(0$^+$)+2$p$ channel($I$=0) gives a minor contribution 
to the 1$^-$($I$=0) phase shift 
and hardly affects the 1$^-$($I$=1) phase shift.
The contribution of the $d$+2$p$ channel is more important 
than that of the other spin-parity states 
of the $^3$He cluster. 
Thus we find that the \{$^3$He(1/2$^+$)+$p$\}+\{$d$+2$p$\} 
calculation including the deuteron  
 ($pn$) and di-proton ($2p$) configurations  
is nearly sufficient to reproduce all the $P$-wave 
phase shifts. 
These results strongly indicate that the low-lying 
resonances of $^4$Li cannot be described adequately in 
the single  configuration of $^3$He(1/2$^+$)+$p$.
This is a sharp contrast to the results in Model C
as is discussed in the following subsection.

In contrast to the $P$-wave 
$^3$He+$p$ scattering phase shifts,  
the $S$-wave phase shifts 
with 0$^+$ and 1$^+$ gain negligible 
contributions from the channels other 
than the main $^3$He(1/2$^+$)+$p$ channel. 
Their phase shifts are shown in Fig.~\ref{fig:g3rsb}. 
A comparison of the $S$-wave and $P$-wave phase shifts 
clearly indicates that the $^3$He+$p$ interaction is 
attractive in the $P$-wave but repulsive in the $S$-wave. The 
calculation suggests that the attractive nature of the $P$-wave 
$^3$He+$p$ resonance cannot be taken into account fully in 
the single $^3$He+$p$ configuration but calls for more complex 
states or distorted configurations which couples with the 
elastic configuration.
 
\subsection{$^3$He+$p$ elastic scattering in Model C}

\begin{figure}[t]
\centerline{\includegraphics[width=12 cm,height=5 cm]{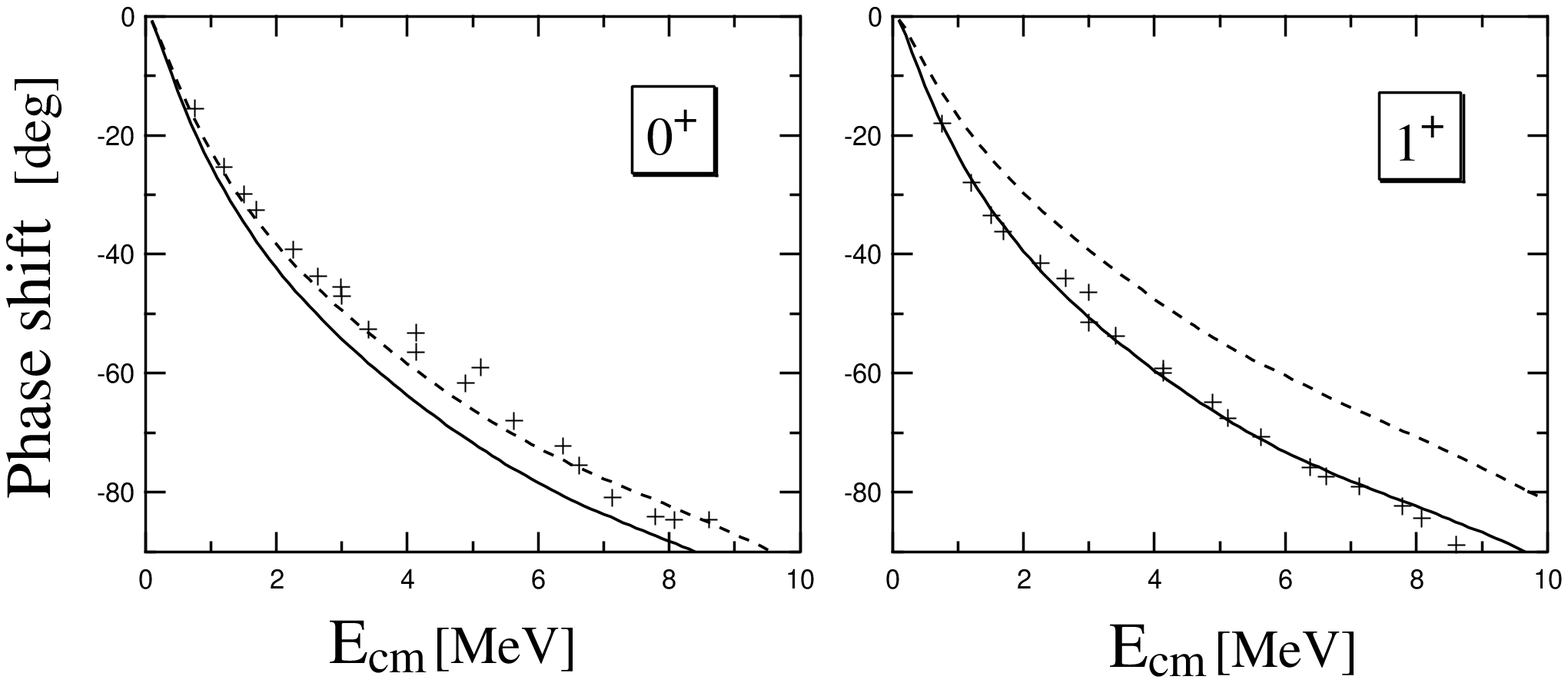}}
\caption{The $^3$He+$p$ $S$-wave
elastic scattering phase shifts calculated by Model C (MN potential).
The wave function of $^3$He is given by
four-range (0$s$)$^3$ h.o. functions.
The solid and dashed lines denote the results of
the $^3$He(1/2$^+$)+$p$ and
\{$^3$He(1/2$^+$)+$p$\} + \{$d$(0$^+$,1$^+$)+2$p$(0$^+$)\}
configurations, respectively.
The crosses denote the experimental data\cite{tombrello65}.} 
\label{fig:mn0sb}
\end{figure}

\begin{figure}[h]
\centerline{\includegraphics[width=10 cm,height=8 cm]{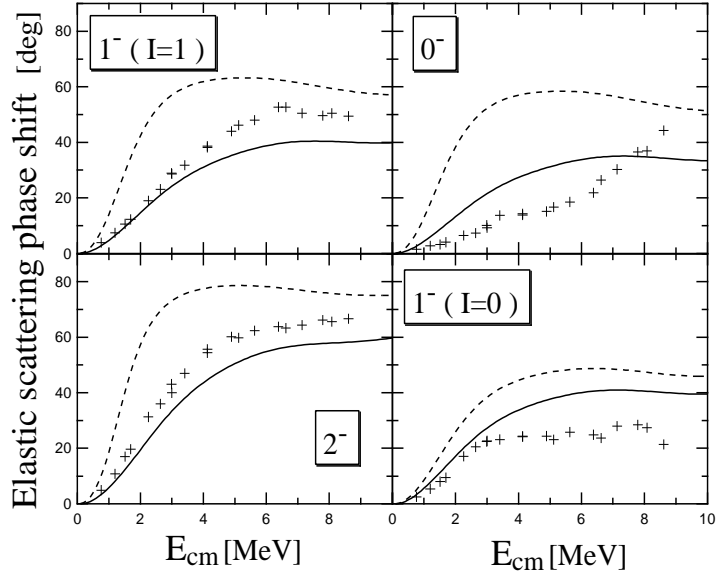}}
\caption{The $^3$He+$p$ $P$-wave elastic scattering phase shifts
calculated by Model C (MN potential). See the caption of 
Fig.~\ref{fig:mn0sb}.}
\label{fig:mn0sa}
\end{figure}

Figures~\ref{fig:mn0sb} and \ref{fig:mn0sa} display 
the $S$- and $P$-wave elastic scattering phase shifts 
which are obtained with Model C
using the MN potential. The $^3$He cluster wave function 
is given by the ($0s$)$^3$ h.o. functions in this calculation. 
The results with the $^3$He+$p$  single configuration 
are shown by the solid lines, and 
those including additionally the $d$+2$p$ channel 
are shown by the dashed lines. 
In this calculation, the $d$ and 2$p$ clusters have only the $S$-wave 
component. 
We see that the $d$+2$p$ channel gives a considerable contribution 
to not only the $P$-wave resonant phase shifts 
but also the $S$-wave non-resonant phase shifts, which 
is in sharp contrast to Model T case shown in 
Figs.~\ref{fig:g3rsa} and ~\ref{fig:g3rsb}.

\begin{figure}[t]
\centerline{\includegraphics[width=10 cm,height=8 cm]{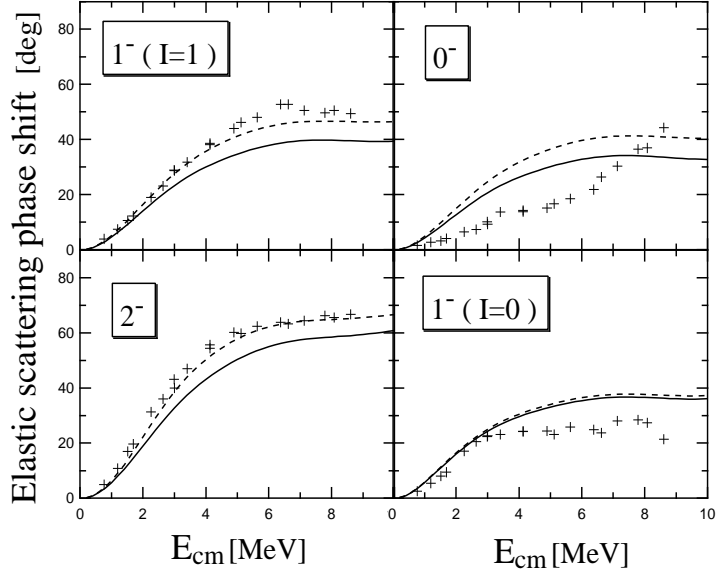}}
\caption{The $^3$He+$p$ $P$-wave elastic scattering phase shifts
calculated by Model C (MN potential).
The wave function of the $^3$He is obtained in 
the $p$+$p$+$n$ three-body model. See the caption of 
Fig.~\ref{fig:mn0sb} for the solid and dashed lines.}
\label{fig:mn3ba}
\end{figure}

The fact that the $d$+2$p$ channel is found to play an important 
role seems to suggest that the distortion of $^3$He has to be 
taken into account. It should be noted, however, 
that the importance of the distortion of the 
clusters may depend on how accurately their wave functions are 
described. In order to examine this issue, 
we repeat the phase shift 
calculation by replacing the $^3$He wave function from 
the simple (0$s$)$^3$ h.o. function  with that of 
the $p$+$p$+$n$ three-body calculation 
as explained in Sect.~\ref{model}. 
The $P$-wave phase shifts which result from this improved 
$^3$He wave function 
are shown in Fig.~\ref{fig:mn3ba}. 
The single channel calculation (solid line) 
of $^3$He+$p$  gives only a minor change on 
both the $S$- and $P$-wave phase shifts, compared to the 
corresponding case of Fig.~\ref{fig:mn0sa}.  
Now let us turn to the effect of including 
the $d$+2$p$ channel on the phase shifts (dashed line). 
In a sharp contrast to the case of Fig.~\ref{fig:mn0sa}, 
we see that 
the calculation using the 
improved $^3$He wave function leads to a significant suppression 
of the additional $d$+2$p$ channel. 
Especially, the contribution in the 1$^-$($I$=0) and two $S$-wave
phase shifts turns out to be negligibly small.

The different role of the  $d$+2$p$ channel mentioned above can be 
explained as follows. 
The two configurations of $^3$He+$p$ and $d$+2$p$ are not 
orthogonal each other and 
have a significant overlap at the short distances of the 
cluster separation. 
The inclusion of the $d$+2$p$ channel plays a role of a distortion effect 
of the $^3$He cluster and modifies the $^3$He wave function indirectly. 
This additional $d$+2$p$ channel has a larger effect 
when the simple 
(0$s$)$^3$ h.o. function is used. Moreover this effect is 
noticeable at low incident energies because 
the $^3$He+$p$ threshold with the (0$s$)$^3$ h.o. function 
is predicted to be about 2.4\,MeV too high compared to the one
with the $p$+$p$+$n$ wave function,  
as seen in Table~\ref{table:1}.
As a whole both of the $S$ and  $P$-wave phase shifts 
are well reproduced by the single $^3$He(1/2$^+$)+$p$ calculation 
if a realistic $^3$He wave function is used. 
To conclude, the 
effect of the cluster distortion strongly depends on whether 
or not the cluster intrinsic wave function is described 
appropriately according to 
the employed effective $N$-$N$ potential.

We have seen that the role of the $d$+2$p$
channel appears quite differently between Model T
and Model C.
In the former case using the realistic potential, 
the $d$+2$p$ channel plays a vital role particularly in the 
$P$-wave resonant phase shifts, 
responding to the complexity due to the tensor force. 
In the case of Model C using the effective potential, 
however, the situation is different. The potential is mainly central 
and induces no complicated angular momentum couplings. Thus 
most of the dynamics are accounted for by the main configuration 
especially when the participating clusters are described realistically,  
and the effects of additional configurations are more or less suppressed. 

A similar suppression by the improvement of the cluster wave 
function was noted in understanding the neutron-halo structure of 
$^6$He in the $\alpha$+$n$+$n$ cluster model. The issue there was the 
role of the additional $t$+$t$ channel\cite{csoto93a,arai99}. 
As was shown in Ref.~\citen{arai99}, the use of the 
simple (0$s$)$^4$ h.o. function for the $\alpha$ particle 
led to the conclusion that the $t$+$t$ channel is really important to 
gain the binding energy of $^6$He, indicating the certain 
deviation from the three-body cluster picture. However, if the 
simple (0$s$)$^4$ h.o. wave function was replaced with the better 
one calculated 
in the $3N$+$N$ two-body model, the effect of 
the $t$+$t$ channel was reduced to a large extent, making it 
possible to maintain the dominant configuration of 
$\alpha$+$n$+$n$. 
This suggests that  we must perform 
the multi-configuration calculation  
paying attention to the cluster intrinsic function 
so as not to overestimate the contribution by the other 
configurations such as the $d$+$2p$ channel in $^4$Li.

\subsection{The ground state of 
$^4$He and $^3$H+$p$ elastic scattering}

In this subsection, we take up two problems 
relevant to  the 0$^+$ state of 
$^4$He, the ground state energy of $^4$He and 
the  $^3$H+$p$ $S$-wave scattering,  
in order to reinforce the arguments 
made in the preceding subsections. The analysis is performed 
in a scheme similar to the $^3$He+$p$ calculation, namely  
using the configurations of 
\{$^3$He(1/2$^+$)+$n$\},  \{$^3$H(1/2$^+$)+$p$\}, 
\{$d$(1$^+$)+$d$(1$^+$)\}, \{$d$(0$^+$)+$d$(0$^+$)\}, and  
\{$2n$(0$^+$)+$2p$(0$^+$)\}.
The intrinsic wave functions of $^3$He, 
$d$, and 2$p$ are the same as in 
the $^3$He+$p$ calculation 
and those of $^3$H and $2n$ are given 
by the same basis sets as those of $^3$He and 2$p$, 
respectively.

The binding energy of $^4$He obtained in Model T 
is $-$24.41MeV for the configuration of  
\{$^3$He+$n$\}+\{$^3$H+$p$\}+\{$d$+$d$\}+\{$2n$+$2p$\}, and 
$-$22.16MeV for \{$^3$He+$n$\}+\{$^3$H+$p$\}, respectively.
The contribution of the $d$+$d$ and 2$n$+2$p$ channels to 
the energy gain is 
2.25MeV. Here the $d$(1$^+$)+$d$(1$^+$) channel gives the 
most important contribution (2.1MeV).
The content of this energy gain is brought about as follows: 
The kinetic energy gives the loss of 8.8MeV, 
while the central and tensor potentials give the gains of  
4.4MeV and 6.9MeV, respectively. 
Thus we see that but for the tensor force, 
the $2N$+$2N$ configuration cannot gain the binding energy. 
Note that the binding energy of the above coupled-channels 
calculation in Model T is only 0.88MeV lower than the more precise 
value, $-$25.29MeV, of the SVM calculation~\cite{suzuki08}. 
Moreover, this difference could be reduced further if 
more extended basis sets are used for $^3$He, $^3$H, and $d$.
In Model C,   
the $u$ parameter of the MN potential is set to $u$=1.0
and the $LS$ term is omitted. 
The resulting energy is  
$-$29.94MeV for the combined configurations of 
\{$^3$He+$n$\}+\{$^3$H+$p$\}+\{$d$+$d$\}+\{$2n$+$2p$\}, and 
$-$29.91 MeV for \{$^3$He+$n$\}+\{$^3$H+$p$\}, respectively
, whereas the SVM energy is $-$29.94MeV. 
Model C thus produces almost fully convergent energy.  
The contribution of the $d$+$d$ and 2$n$+2$p$ channels 
is only 0.03MeV, which is much smaller than in Model T.

\begin{figure}[t]
\centerline{\includegraphics[width=12 cm,height=5 cm]{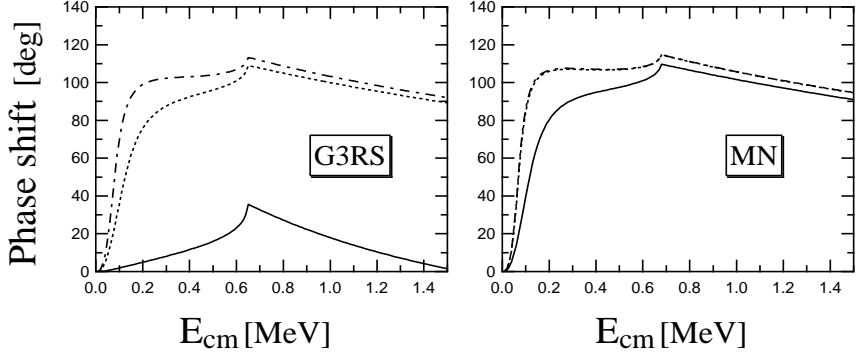}}
\caption{The $^3$H+$p$ $S$-wave 0$^+$ elastic scattering phase shift
by Model T with the G3RS potential(the left panel) 
and by Model C with the MN potential(the right panel).
The lines denote the results obtained 
including the following configurations:
Solid  \{$^3$H+$p$\}+\{$^3$He+$n$\};
\, Dotted \{$^3$H+$p$\}+\{$^3$He+$n$\}+\{d(1$^+$)+d(1$^+$)\} ;
\, Dash-Dotted
 \{$^3$H+$p$\}+\{$^3$He+$n$\}+\{d(1$^+$)+d(1$^+$)\}+
\{d(0$^+$)+d(0$^+$)\}+ \{2n(0$^+$)+2p(0$^+$)\}.}
\label{fig:t+p}
\end{figure}

Now we come to the $^3$H(1/2$^+$)+$p$ $S$-wave 0$^+$ 
elastic scattering phase shift.
Figure~\ref{fig:t+p} compares the phase shifts between  
Model T and Model C.  
The solid, dotted, and dash-dotted lines 
denote the results using 
the configuration of \{$^3$H+$p$\}+\{$^3$He+$n$\},
\{$^3$H+$p$\}+\{$^3$He+$n$\}+\{$d$(1$^+$)+$d$(1$^+$)\}, 
and \{$^3$H+$p$\}+\{$^3$He+$n$\}+\{$d$+$d$\}+\{$2n$+$2n$\},
respectively. The contribution of the $d$+$d$ 
channel is very different, depending on the model. 
It is very large in Model T but 
much less significant in Model C.

In order to discuss how much 
the $d$+$d$ configuration is different from the $^3$H+$p$ 
configuration between Models T and C,  
we calculate the following overlap 
\begin{equation}
\left< {\cal A}\Big\{ 
[[ \Phi^{^3{\rm H}}_{1/2^+} \Phi^p_{1/2^+}]_{I=0} \; 
\Gamma_{\ell=0}(\nu,\mbox{\boldmath$\rho$})]_{J=0^+} \Big\}
\left| 
{\cal A}\Big\{ [[\Phi^d_{1^+} \Phi^d_{1^+} ]_{I'=0} \; 
\Gamma_{\ell'=0}(\nu',\mbox{\boldmath$\rho$}')]_{J=0^+} \Big\} 
\right> \right. , 
\label{eqn:b1}
\end{equation}
where the wave functions in bra and ket are both normalized to 
unity.
We take a single Gaussian basis function with a 
common parameter $\nu=\nu'$. 
The overlaps for $b$(=$1/\sqrt{\nu})$=1.0, 2.0, 3.0, and 4.0\,fm 
are 0.53, 0.90, 0.96, and 0.93 in Model T, respectively, while they 
are 0.60, 0.92, 0.97, and 0.93 in Model C. 
The $d$+$d$ configuration has a smaller 
overlap with the $^3$H+$p$ configuration in Model T than in Model C 
when the two $d$ clusters come closer than $b<1$\,fm. 

We understand 
the different contribution of the $d$+$d$ channel
between Model T and Model C as follows.
In the bound or resonance state where 
the four nucleons are localized in the interaction region 
through the attractive interaction of the clusters,  
various types of correlations like $3N$+$N$ and $2N$+$2N$
are equally important.
If the overlap between the different configurations is large 
and the structure of the state is relatively simple, 
a particular channel can accommodate most of the indispensable 
configurations fairly well as in Model C. 
However, once the tensor force is explicitly 
taken into account and the higher partial waves 
are included in the cluster intrinsic 
wave functions, the overlap between the different 
configurations becomes smaller at the 
short cluster relative distance
and in addition the structure of the bound or resonance 
state becomes more complicated.
As a result, the state cannot be well described  
with a single configuration. 
Contrary to the bound or resonance state, 
the non-resonant state in the $^3$He+$p$ $S$-wave 
scattering is well approximated with the single configuration
because the two clusters feel a repulsive interaction 
in the scattering and the chance of coupling with the other 
channel becomes small.

\section{Summary}

We have calculated the $^3$He+$p$ $S$- and $P$-wave 
elastic scattering phase shifts in two different 
microscopic cluster models, Model T and Model C.  
The $s$-shell cluster intrinsic function includes the $D$-waves 
through the tensor force in Model T,
while it is described with 
only the $S$-wave Gaussian function in Model C. 
These models have also been applied to 
the 0$^+$ ground state of $^4$He 
and the $^3$H+$p$ $S$-wave elastic scattering 
phase shift in order to elucidate the 
role of different cluster channels.

We have found that, 
in Model T using a realistic nucleon-nucleon interaction, 
the inclusion of the $d$+2$p$ channel is very important 
to reproduce the $^3$He+$p$ $P$-wave resonant phase shifts, whereas 
the single $^3$He(1/2$^+$)+$p$ channel alone can reproduce the 
$S$-wave non-resonant phase shifts fairly well.  

In contrast to the realistic interaction case, 
in Model C where an effective interaction is used,  
the role of the $d$+2$p$ channel depends on how realistically  
the $^3$He wave function is described.
If it is given by 
the simple (0$s$)$^3$ harmonic-oscillator function, 
the $d$+2$p$ channel has contributed 
significantly to both the $P$- and $S$-wave phase shifts.
This is because the distortion effect of the $^3$He cluster 
cannot be taken into account sufficiently by the simple (0$s$)$^3$ 
harmonic-oscillator 
function and the $d$+2$p$ channel indirectly modifies 
the $^3$He cluster intrinsic wave function.
However, if it is improved with the $p$+$p$+$n$ three-body
wave function, we have confirmed that the contribution 
of the $d$+2$p$ channel is greatly suppressed even in the 
$P$-wave resonant phase shifts. 
In comparison with the model T, 
these results suggest that
the  $d$+$2p$ channel is renormalized mostly 
as the the $^3$He(1/2$^+$)+$p$ channel in the resonance region
in the model C.

We have obtained similar results for both the binding energy
of $^4$He and the $^3$H+$p$ $S$-wave elastic scattering phase shift.
The $d$+$d$ channel has a significant contribution in Model
T, while it plays a minor role in Model C.
We have shown that in Model T the $d$+$d$ channel 
is important to improve the short-range behavior of the 
four-nucleon wave function.
Without the tensor force, the energy gain due to 
the $d$+$d$ channel caused by the central potential is much smaller 
than the energy loss of the kinetic energy, resulting in the minor 
contribution of the $d$+$d$ channel.

\section{Acknowledgements}
This work presents research results of Bilateral Joint Research 
Projects of the JSPS (Japan) and the FNRS (Belgium).

%

\end{document}